# Microscopic Filamentary Theory of Electronic Intermediate Phases

J. C. Phillips

Dept. of Physics and Astronomy, Rutgers University, Piscataway, N. J., 08854-8019

**Abstract**

The self-organized dopant percolative filamentary model, entirely orbital in character (no spins), is strongly influenced by background (in)homogeneities. In the high temperature superconductive (HTSC) cuprates, pseudogap regions produce novel percolative features in phase diagrams. The model is especially successful in describing the appearance of giant magnetic vortex "precursive" effects at temperatures ~2 $T_c$ far above the superconductive transition temperature $T_c$.

High temperature cuprate superconductors [1] exhibit many counter-intuitive properties, as revealed in a wide variety of experiments. Conventional theoretical methods encounter serious difficulties in explaining these anomalies microscopically, partly because of the complexity of the unit cells, partly because superconductivity depends entirely on doping (the undoped parent compounds are insulating), and partly because the materials are intrinsically inhomogeneous, and exhibit two energy gaps, a superconductive gap and a semiconductive pseudogap. The two gaps create a patchy nanoscale pattern; disentangling the two gaps has so far proved difficult even for large scale, high resolution scanning tunneling microscope studies [2].

Almost all the general tools of condensed matter theory (mainly band theory and toy models) have been able to describe one or another aspects of these complex materials, but in almost all cases a mean-field approximation is used to reduce exponential complexity to the polynomial level. The neglect of strong nanoscale disorder simplifies polynomial mean-field models, but it also makes it necessary usually to assume what is later "proved". There are two key features of cuprates that should be derived, not assumed: the existence of an intermediate phase, sandwiched between conventional insulating and metallic (Fermi liquid) phases, and the presence of gigantic precursive effects, extending as high as twice the superconductive transition temperature $T_c$. in temperature in transport properties, as high as 1.5 eV in optical experiments, and as high as 20-30 eV in photoemission experiments [3]. Both key features emerge naturally in non-analytic topological models that model the electrically active dopants as a self-organized, *space-filling* network ("pearls on a string") [4-11].

Here I extend this amazingly successful model to two very large precursive magnetic effects: the giant Nernst effect in the thermomagnetic power [12], and field-enhanced diamagnetism, with composition dependence parallel to that of the Nernst effect [13]. Because of the glassy nature of the dopant network, special tools are needed to analyze its quantum coherence. The static configuration space defined by the filamentary dopant network is special in itself, but this problem is by now quite well understood. Following Boolchand's discovery of the intermediate phase in glass networks and supercooled liquids [5], several numerical simulations have shown the existence of this specially connected (linear topological dimensionality, but not mathematically one-dimensional) phase [6-11]. It is a novel phase, dependent on long-range interactions (elastic or electrical), and it is quite generally absent from mean-field theories [14], which was discovered later in the context of computability of Boolean algebras [15].

The semiclassical Bohr quantum model, formalized as wavelets, is an appropriate tool for including coherence effects and efficiently constructing electrical conductivity paths in heterogeneous materials at low temperatures [16]. It can also be combined with Fourier grid methods to speed up considerably simulations of wave packet propagation in multiscale systems [17] analogous to glassy networks embedded in crystalline hosts. Using such methods, one can in principle construct both one-electron generalizations of periodic crystalline states, and many-electron generalizations of BCS Cooper-paired states to the case of strongly disordered dopant networks. The present discussion identifies many important qualitative points without carrying out numerical wavelet simulations, but it is gratifying to note that once the dopant configurations have been properly characterized topologically, one could carry out either classical [6-11] or quantum [16-17] simulations to elucidate properties in more detail.

The simplest case is a homogeneous medium with randomly distributed dopants; it is illustrated in Fig. 1(a). Although theories often assume that impurities are randomly distributed, preparing a sample of even a simple material such as Si:P with randomly distributed substitutional P dopants is not easy: in fact, it can be done best by neutron transmutation doping (otherwise there is enough clustering during growth due to short-range chemical forces to spoil the threshold of the metal-insulator transition). This subject has a long and interesting history; originally Mott assumed that the transition would be first order, while classical percolation predicts the threshold conductivity to be linear in $n - n_c$, the excess dopant density. Neither prediction was correct for ideally randomly distributed dopants. Classical continuum methods succeed in a critical region where mean field theory is valid [18], but this region is 100 times narrower than the intermediate phase, which is dominated by quantum percolation [19]. In other words, even in the simple "random" case, quantum effects and self-organization into filaments are the dominant effects, driven by long-range Coulomb interactions.

Because perovskite and especially cuprate pseudoperovskite crystals are anomalously soft, nanoscale disorder and mixing of metallic and semiconductive appear to be characteristic of the cuprates. Because the parent compounds are antiferromagnetic, it has often been assumed that the pseudogap regions are AF remnants, but by now the experimental evidence strongly suggests that the pseudogap patches contain localized charge density waves, probably associated with a 2x2 checkerboard reconstruction. (Of course, the latter is much less likely to destroy Cooper pairs by spin scattering.) It now seems quite likely that there are also two kinds of dopants, for example, two kinds of interstitial oxygen $O_x$ atoms, as in $Bi_2Sr_2CaCu_2O_{8+x}$ (BSCCO): electrically active $O_x^-$ pinned to $E_F$, and $O_x^{2-}$, with an energy level ~ -0.9 eV, concentrated mainly in pseudogap regions [20]. In any case, the pseudogap regions represent barriers in a maze around which filamentary paths centered on electrically active dopants must percolate.

As with all disordered percolative structures, local structure is characterized by its longitudinal and transverse features (Fig. 1(b)). In the conducting regime there are local curvelinear channels that are longitudinally open, but transversely confined, with a variable spacing between the pseudogap confinement walls. Within the tetragonal framework of conductive $CuO_2$ planes one can distinguish two principal directions, the (10) ones parallel to CuO rows, and the (11), angularly between the (10) and (01) CuO rows. Electron-phonon scattering is strongest parallel to CuO rows, so one expects that mean free paths will be longer in the (11) directions than in the (10) directions. In other words, the Fourier grid of the host lattice lifts the planar angular degeneracy.

At high temperatures the dopants will be strongly disordered, but at the annealing temperatures used to optimize sample properties they are still strongly mobile, especially if they are interstitial oxygen $O_x^-$ atoms, as in $Bi_2Sr_2CaCu_2O_{8+x}$ (BSCCO). (Even the $O_x$ atoms in the $CuO_x$ plane of $YBa_2Cu_3O_{6+x}$: are not all nicely lined up in perfect rows.) These mobile electrically active dopants adopt *glassy* (non-crystalline) configurations that maximize the dynamical dielectric screening of fluctuating internal electric fields in these ionic oxides. (Note that if the channels were rectilinear, as they are in the superlattice 1/8 phase of LSCO, the dopant sites would have a strong tendency towards periodicity. This would stiffen the dopants, greatly reduce the strength of attractive electron-phonon interactions, and would almost certainly quench superconductivity, as observed.) In Fig. 1(c) this leads at low densities, just below the metal-insulator transition (MIT), to the formation of curvilinear filamentary segments. These segments have been observed in ARPES experiments [3] in the benchmark Bednorz-Mueller alloy family $La_{2-x}Sr_xCuO_4$ (LSCO), where the MIT occurs at $x_c = 0.06$, even at x = 0.03. The intensity of the energy distribution at $E = E_F$ is proportional to the quasi-particle strength $Z(\mathbf{k},E_F)$; as x increases, $Z(\mathbf{k},E_F) > 0$ first appears in the nodal $\mathbf{k} = (\pi,\pi)$ gap direction, and then spreads smoothly over a wider angle as x increases. This is what one expects from coherent screening, as mean free paths are longer in the (11) directions than in the (10) directions.

The coherence of each filament requires that the dopant states overlap enough longitudinally to be metallic, but that they do not overlap enough transversely to form Fermi liquid patches; the specific conductivity of the latter is smaller than that of the filaments, because the scattering is smaller in the latter, being tangentially restricted to one dimension. Thus the number of curvilinear filaments/channel is limited by bottlenecks in the lateral confinement dimension d($\xi$). As an example, the channel in Fig. 1(d) locally accommodates one longer and two shorter nearly linear filaments.

When there are large variations in the channel aspect ratio (D/d, Fig.1(b)), some of the dielectric screening filaments will form closed loops that support diamagnetic currents (Fig. 2(a)). The average area/atom of these loops would appear to be similar to that of a nearly linear segment that crosses a bottleneck between two wider regions. All other things being equal, static dielectric screening energies favor nearly linear segments over closed loops of the same length. To see this, one can look at the screening diagrams. In Fig.2 (b) one has two contacting circular dopants. The screening energy is increased over two separated circular dopants, because the potential strength of the two potentials spreads out beyond the two circles (dashed line)and the screening charges overlap.. The elliptical details of how this occurs are unimportant, because at this stage we are studying only polynomial (essentially quadratic) effects. When we add a third dopant, however, we are starting a series that will eventually lead to an entropically exponentially small region of configuration space. One sees immediately that the nearly linear segment (c) screens a (modulated) dielectric strip, while incipient loops have a smaller static dielectric screening energy.

What this means is that at low dopant densities x ~ $x_c$, linear segments are energetically favored over closed loops of the same length. However, as x increases towards optimal doping x = $x_0$, curvilinear strips fill most of the bottleneck space available. Dynamical screening power depends on the local conductivity, and with decreasing T the loop conductivity can be larger than that of a linear segment of equal length, because of scattering at the ends of the latter. Thus both dynamical and packing factors will cause dopants to form closed loops in the empty space of "ponds". In the presence of crossed electric (or thermal gradient) and magnetic fields, the vorticity of carriers in these loops can be activated to jump across the bottlenecks to loops in adjacent "ponds". Because the hopping distance is large, and the pinning of the vortices by the soft host lattice is weak, this mechanism will inevitably lead to a giant Nernst effect in the thermomagnetic power. This effect has been observed [12] in both BSCCO and LSCO, and it peaks near x = 0.11, in other words, well above the MIT at x = $x_c$ = 0.06, yet well below optimal doping x = $x_0$ = 0.16, near x = $(x_c + x_0)/2$, which would be the expected value for a plausible channel aspect ratio of 2. The onset temperatures at optimal doping $x_0$ are 150K (BSCCO) and 90K (LSCO), well above $T_c$, and well below pseudogap temperatures.

From the "tilted hill" magnetic field dependence that is observed both above and below $T_c$, the authors conclude [12] that the vortices are continuously present in the normal as well as the superconductive state. This agrees perfectly with the present topological analysis, as does their observation that the vortices are inherently connected to pseudogaps. (In fact, the topological dielectric screening model is even stronger than their model. They assume that the energy for creating vortices in a magnetic field is smaller ("cheap") than in the homogeneous case (no pseudogaps), but at low temperatures the dynamical loops in the present model are *native* to the self-organized dopant network [zero, or negative, formation energy].) The data present very serious difficulties for the many other theories they discuss, as those theories are all based on homogeneous mean-field lattice models without specific adaptive dopant configurations. These theories involve different kinds of additional assumptions (including analogies with elementary particles and liquid crystals [12]), and they do not predict the dopant composition dependence, with its maximum near x = $(x_c + x_0)/2$. The presence of native diamagnetic loops also predicts field-enhanced diamagnetism, with composition dependence parallel to that of the Nernst effect, as observed [13]. Note that both vortex hopping and field loop enhancement assume that there is some "free area" available to vortex dynamics. This "free area" disappears as x → $x_0$, where both effects become smaller.

Given the remarkable softness of the cuprates [20], one can expect to find evidence in their structures for filamentary formation – providing one is successful in searching the very large (~100,000 papers!) cuprate literature. Several such recent examples have already been noted [4]. However, the most spectacular data found so far are in an old paper [21], which reported data on the average kinetic energy <E>(T) of Cu ions in BSCCO measured by resonant neutron absorption spectroscopy; these are reproduced for the reader's convenience in Fig. 3. The results for optimally doped BSCCO (<R> = 2.14) show a much steeper linear <E>(T) for high T > 140K than for undoped LCO (<R> = 2.29); as expected, the smaller <R> softens BSCCO and increases $T_c$. . (Thus the onset T is about the same for the Nernst effect and Cu kinetic energy arrest.) The most interesting behavior, however, is the softening between onset at 140K and $T_c$. This precursive effect becomes very steep near $T_c$, and it can be interpreted as a mean-field transition that "freezes" filamentary ends. (The strain energy of a filamentary segment is harmonic, and so is proportional to (the length) $l^2$; increasing the length by $\Delta l$ increases the energy by $l\Delta l$, so that $l$ itself is a mean field for filamentary strain energies.) The dopant vortices themselves could be formed at higher temperatures where pseudogaps first appear [22].

In conclusion, it appears that the counter-intuitive properties of high temperature cuprate superconductors are inextricably linked to the exponential (non-polynomial) complexity of their aperiodic off-lattice dopant ("pearls on a string") networks. Modern computer science [14,15] recognizes that such complexities are not easily handled using the Newtonian polynomial methods that have often been used to discuss liquid crystals (Landau continuum theory) or elementary particles (t-J lattice models). However, just as for network and even hard sphere glasses [23], these remarkable properties appear to be well described by Lagrangian topological theories of connectivity in appropriate configuration spaces [24]. This means that the complexity of the cuprates is ideally suited to topological modeling.

## Figure Captions

Fig. 1. (a) Randomly distributed dopants (black dots) in a homogeneous background. Ionic fluctuations are screened dielectrically in regions here idealized as green circles. (b) Channel geometry formed by pseudogap regions: the black dots are $O_x^-$, while the red rectangles are $O_x^{2-}$. (c) Even at low doping densities it is energetically advantageous for the electrically active dopants to form linear segments (upper chain). (d) More than one linear segment can form in wide regions of a channel; when the segments lengthen, narrow regions act as bottlenecks, limiting the number of percolating segments.

Fig. 2. (a) When there is free area in wide regions which are shorter than shown in Fig. 1(d), closed loops (hydrodynamic eddies) are energetically favored when all the area available to longer filaments has been filled. (b) Adjacent non-overlapping dopant pairs screen larger areas. Linear segments (c) screen larger areas than curved segments (d). In these figures, double arrows in each circle represent enhancement of screening along enhanced polarizability axes due to charge overlap (not shown).

Fig. 3. $<E>(K)$ of Cu ions [21]. The dashed line is a scaled calculation for a parent (undoped) compound, LCO. Note that doping linearizes and softens $<E>(K)$ for $T > 145$ K, and introduces an abrupt stiffening for $T < 140K$, which then shows a "phase transition" as $T \to T_c$.

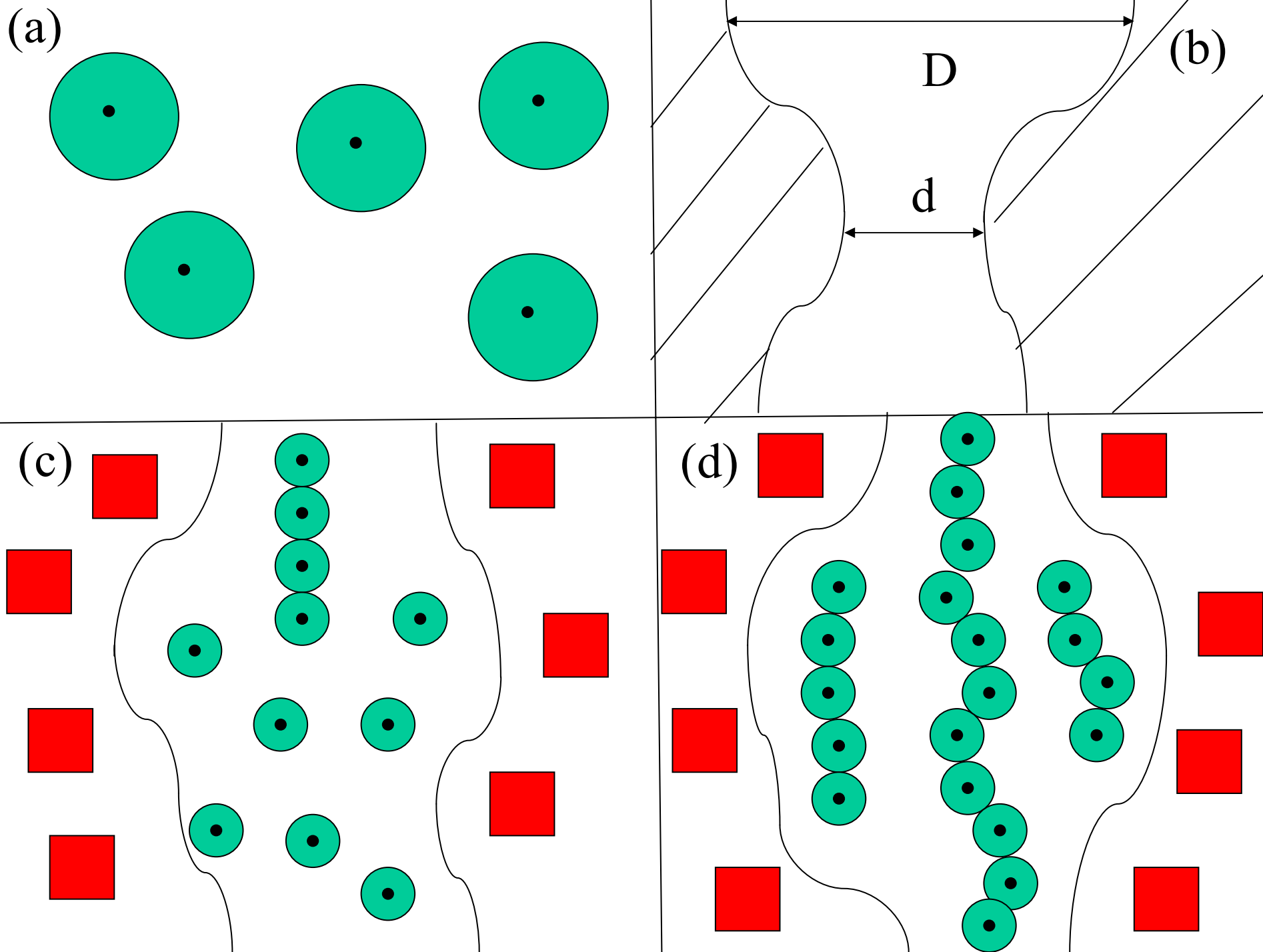

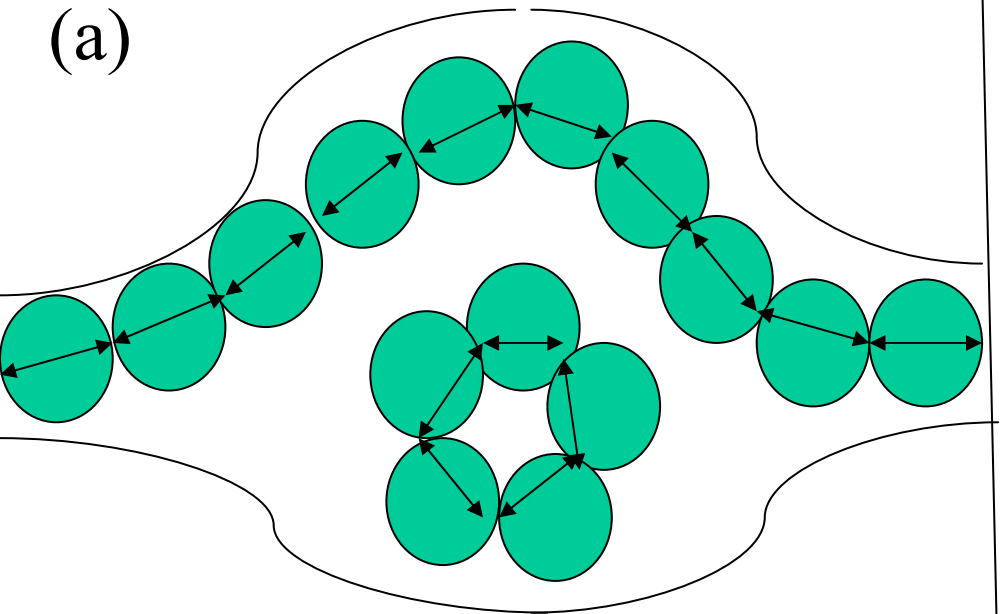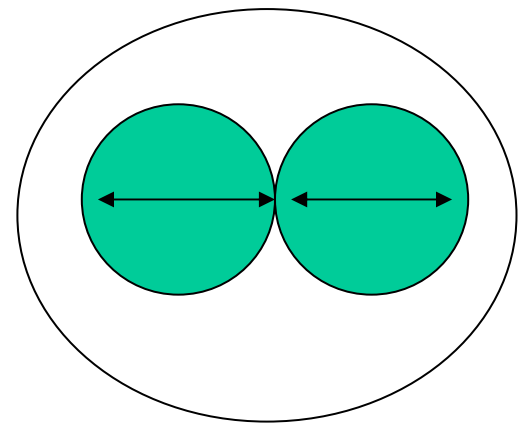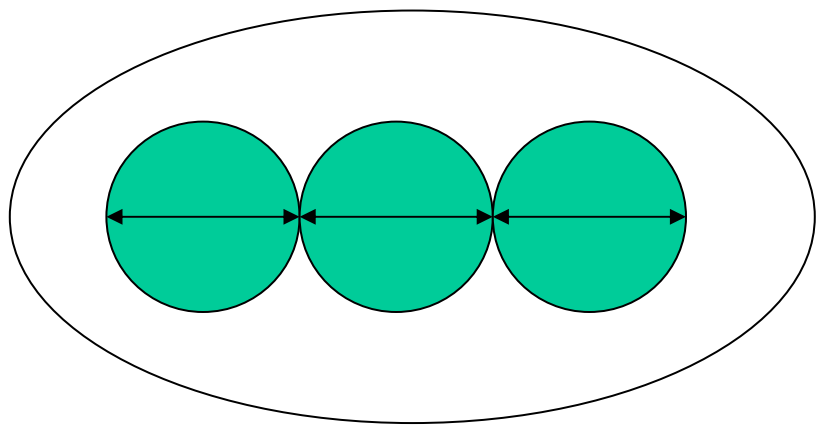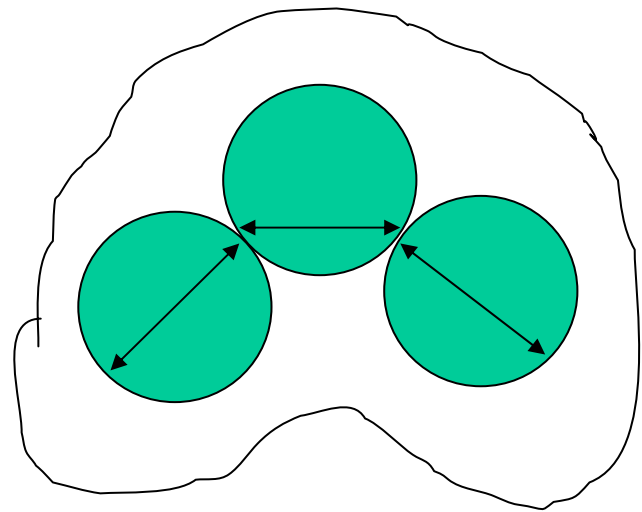

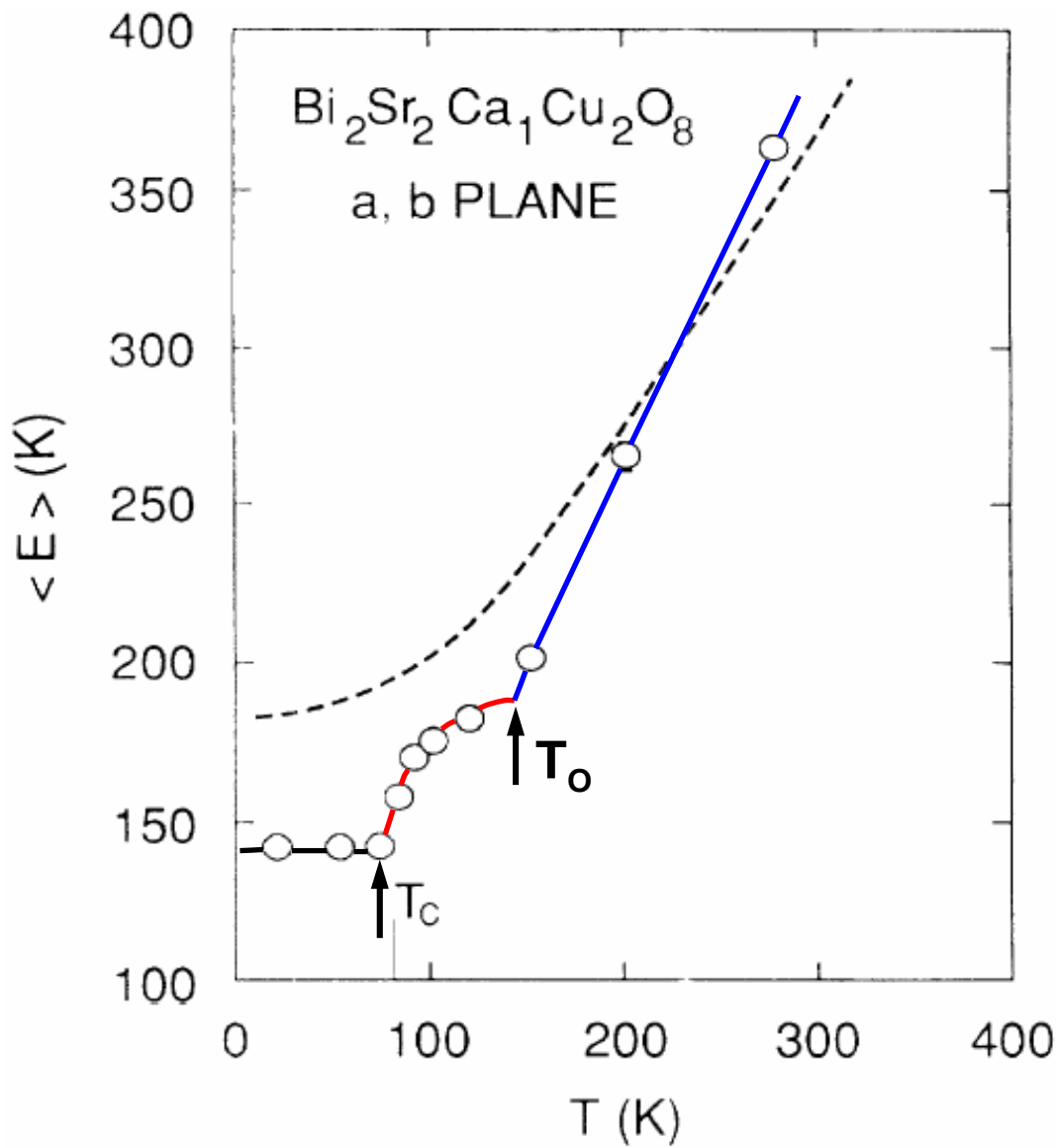